\documentclass{ws-rv975x65}
\usepackage{ws-rv-van}             
\makeindex
\usepackage{graphicx}
\usepackage{color}

\begin{document}

\chapter{Anderson localization \emph{vs.} Mott-Hubbard metal-insulator transition
in disordered, interacting lattice fermion systems\label{ch1}}

\author[K. Byczuk, W. Hofstetter, and D. Vollhardt]{Krzysztof Byczuk,$^{1,2}$ Walter Hofstetter,$^{3}$  and Dieter Vollhardt$^2$
}

\address{$^1$ Institute of Theoretical Physics, University of Warsaw, ul. Ho\.za 69,\\
PL-00-681 Warszawa, Poland\\
$^2$Theoretical Physics III, Center for Electronic Correlations and Magnetism,
Institute for Physics, University of Augsburg, D-86135 Augsburg, Germany\\
$^3$Institut f\"ur Theoretische Physik, Johann Wolfgang Goethe-Universit\"at,\\
60438 Frankfurt/Main, Germany
}

\begin{abstract}
We review recent progress in our  theoretical understanding of
strongly correlated fermion systems in the presence of
disorder. Results were obtained by the application of a powerful
nonperturbative approach, the Dynamical Mean-Field Theory (DMFT), to
interacting disordered lattice fermions. In particular, we demonstrate
that DMFT combined with geometric averaging over disorder can capture
Anderson localization and Mott insulating phases on the level of
one-particle correlation functions. Results are presented for the
ground-state phase diagram of the Anderson-Hubbard model at half
filling, both in the paramagnetic phase and in the presence of
antiferromagnetic order.  We find a new antiferromagnetic metal which
is  stabilized  by disorder. Possible realizations of these quantum
phases with ultracold fermions in optical lattices are discussed.
\end{abstract}

\body

\section{Introduction}\label{sec1.1}

In non-interacting quantum systems with disorder, e.g., in the
presence of randomly distributed impurities, wavefunctions can either
be spatially extended or localized. Until 1958 is was believed that a
localized state corresponds to a bound state of an electron at the
impurity. By contrast, in his landmark paper of 1958 Anderson
\cite{Anderson58} predicted that disorder can lead to quite a
different type of localized state now referred to as ``Anderson
localized state''. To understand its physical origin it should be
noted that if a particle is inserted into a disordered system it will
start to spread.   As a consequence the wave is backscattered by the
impurities, leading to characteristic ``weak localization'' effects
\cite{Abrahams79,Gorkov79,Lee85}. The multiple scattering of the
electronic wave can enhance these perturbative effects to such a
degree that the electrons become spatially localized; for reviews see
\cite{Lee85,VW92,Lagendijk09}. In this case there is a finite
probability for an electron to return to the point where it was
inserted. If states are extended, this probability is zero. So, in
contrast to localized states bound at an impurity, Anderson localized
states are confined to a region of space due to coherent
backscattering from randomly distributed impurities.

In the thermodynamic limit the excitation spectrum determined from the
resolvent of the one-particle system or the one-particle Green
functions is very different for extended and localized states.  The
one-particle Green function describing an extended state has a branch
cut on the real axis, and the spectrum of the Hamiltonian is
continuous. By contrast, the Green function for a localized state has
discrete poles located infinitely close to the real axis, which
implies a discrete point spectrum of the Hamiltonian. In particular,
the point-like spectrum of an Anderson localized state is dense.

In the presence of interactions between the electrons the same
classification of (approximate) eigenstates  may, in principle, be
used. Namely, if the one-particle Green function of the interacting
system has a branch cut at some energies, the states at those energies
are extended. If the Green function has discrete, separate poles the
corresponding states are bound states, and if the poles are discrete
and lie dense the states are Anderson localized.Since one-particle
wave functions  are not defined in a many-body system they cannot be
employed to describe the localization properties of the
system. Instead the reduced one-particle \emph{density matrix}, or the
one-particle Green function $G({\bf r}-{\bf r'})$ in position
representation, may be employed. For localized states these quantities
approach zero for $|{\bf r}-{\bf r'}|\rightarrow \infty$. For extended
states their amplitude only fluctuates very weakly, i.e., of the order
$1/V$, where $V$ is the volume of the system.

In the following we are interested in the question how states of
many-body systems change when the interaction and/or the disorder are
varied. In general, the very notion of a metal or an insulator is
related to the properties of two-particle Green functions, e.g., the
current- and density-correlation functions. There exist different
approaches to study the disappearance of a diffusion pole at the
metal-insulator transition, and correspondingly, the vanishing of the
DC conductivity in the thermodynamic limit
\cite{Lee85,VW92,Mott90,Belitz94}. On physical grounds it is very
plausible to expect that the presence of Anderson localized states
with dense, point-like spectrum at the Fermi level, discussed above in
terms of one-particle Green function, implies zero
conductivity. Mathematical proofs of this conjecture exist only for
specific models and in limiting cases.\cite{Kirsch07} Indeed, it is
usually assumed that the presence of Anderson localized states at the
Fermi level implies the system to be an Anderson insulator, at least
in the non-interacting case. This is also our line of approach which
will be reviewed in this article.

The paper is structured as follows:
In Section 2 we review general aspects of the interplay between interactions and disorder in lattice fermion systems. In particular, we discuss the important question concerning the appropriate average over the disorder, and describe the new developments in the field of cold atoms in optical lattices which will make it possible in the future to investigate disordered, interacting lattices fermions with unprecedented control over the parameters. The models of correlated fermions with disorder are introduced in Section 3, followed by an introduction into the dynamical mean-field theory (DMFT) (Section 4) and a more detailed discussion of arithmetic vs. geometric averaging over the disorder (Section 5). In Section 6 the DMFT self-consistency conditions for disordered systems are introduced. After having defined the characteristic quantities which help us to identify the different phases of the Anderson-Hubbard Hamiltonian (Section 7), the results for the ground state phase diagram at half filling are reviewed (Section 8). In Section 9 the results are summarized.

\section{Interplay between interactions and disorder in lattice fermion systems}
\subsection{Interactions \emph{vs.} disorder}

The properties of solids are strongly influenced by the interaction
between the electrons and the presence of disorder
\cite{Lee85,Mott90,Belitz94}. Namely, Coulomb correlations and
randomness are both driving forces behind metal--insulator transitions
(MITs) which involve the localization and delocalization of
particles. While the electronic repulsion may lead to a Mott-Hubbard
MIT \cite{Mott49,Mott90}, the coherent backscattering  of
non-interacting particles from randomly distributed impurities can
cause Anderson localization  \cite{Anderson58,Abrahams79}.

Since electronic interactions and disorder can both (and separately)
induce a MIT, one might expect their simultaneous presence to be
even more effective in localizing electrons. However, this is not
necessarily so. For example, weak disorder is able to weaken the
effect of correlations since it redistributes states into the Mott gap
and may thus turn an insulator into a (bad) metal. Furthermore,
short-range interactions lead to a transfer of spectral weight into
the Hubbard subbands whereby the total band-width and thus the
critical disorder strength for the Anderson MIT increases, implying a
reduction of the effective disorder strength. Hence the interplay
between disorder and interactions leads to subtle many-body effects
\cite{Lee85,Belitz94,Lohneysen00,Kravchenko04,Finkelshtein83,Castellani84,Logan93,Shepelyansky94,Denteneer01},
which pose fundamental challenges for theory and experiment not only
in condensed matter physics
\cite{Mott90,Lee85,Altshuler85,Belitz94,Abrahams01}, but most recently
also in the field of cold atoms in optical lattices
\cite{Lewenstein07,Bloch08,Fallani07,Billy08,Roati08,White09,Aspect09,Lewenstein10}.
Indeed, ultracold gases have quickly developed into a fascinating new
laboratory for quantum many-body
physics.\cite{Lewenstein07,Bloch08,Jaksch98,Greiner02,HTC,Koehl05,bose-fermi,Mott}
 A major advantage of cold atoms in optical lattices is the high degree
of controllability of the interaction and the disorder strength,
thereby allowing a detailed verification of theoretical
predictions. The concepts, models, and techniques for their
solution to be discussed in this paper equally apply to electronic
systems and cold fermionic atoms in optical lattices. In the following
we will therefore refer generally to the investigation of ``correlated
lattice fermion systems''.

\subsubsection{Average over disorder}

In general the theoretical investigation of disordered systems
requires the use of probability distribution functions (PDFs) for the
random quantities of interest. Indeed, in physical or statistical
problems one is usually interested in ``typical'' values of random
quantities which are  mathematically given by the most probable value
of the PDF \cite{definition1}. However, in many cases the complete PDF
is not known, i.e., only limited information about the system provided
by certain averages (moments or cumulants) is available. In this
situation  it is very important to choose the most informative average
of a random variable. For example, if the PDF of a random variable has
a single peak and fast decaying tails the typical value of the random
quantity is well estimated by its first moment, known as the
\emph{arithmetic} average (or arithmetic mean). But there are many
examples, e.g., from astronomy, the physics of glasses or networks,
economy, sociology, biology or geology, where the knowledge of the
arithmetic average is insufficient since the PDF is so broad that its
characterization requires infinitely many
moments.\cite{Limpert01,lognormal} Such systems are called
non-self-averaging. One example is Anderson localization: when a
disordered system is close to the Anderson MIT \cite{Anderson58}, most
electronic quantities fluctuate strongly and the corresponding PDFs
possess long tails which can be described by a log-normal distribution \cite{Mirlin94,Janssen98,Schubert03,Mirlin08,Science10,Schubert10}. This is well
illustrated by the local density of states (LDOS) of the disordered system. Most recently it was shown for various lattices in  dimensions $d=2$
and $3$  that the system-size dependence of the LDOS distribution is
an unambigous sign of anderson localization, and that the distribution of the LDOS of disordered electrons  agrees with a
log-normal distribution over up to ten orders of magnitude\cite{Schubert10}. Therefore it is not surprising that the
arithmetic mean of this random one-particle quantity does not resemble
its  typical value at all. In particular, it is non-critical at the
Anderson transition\cite{Lloyd69,Thouless74,Wegner81} and hence cannot
help to detect the localization transition. By contrast the
\emph{geometric} mean \cite{Limpert01,lognormal,Montroll83,Romeo03} of the LDOS, which represents the most probable (``typical'') value of a log-normal distribution, is the appropriate average in this case. It vanishes at a critical strength of the disorder and hence
provides  an explicit criterion for Anderson localization in disordered systems\cite{Anderson58,Schubert03,Schubert10}, even in the presence of interactions
\cite{Dobrosavljevic97,Dobrosavljevic03}.

\subsubsection{Dynamical mean-field approach to disordered systems}

In general MITs occur at intermediate values of the interaction and/or
disorder. Theories of MITs driven by interaction and disorder
therefore need to be non-perturbative. Usually they cannot be solved
analytically, and require numerical methods or self-consistent
approximations. A reliable approximate method for the investigation of
lattice fermions with a local interaction is provided by dynamical
mean--field theory (DMFT),\cite{Metzner89,Georges96,PT} where the
local single--particle Green function is determined
self-consistently. If in this approach the effect of local disorder is
taken into account through the arithmetic mean of the LDOS
\cite{Ulmke95} one obtains, in the absence of interactions, the
well-known coherent potential approximation (CPA). \cite{Vlaming92}
However, the CPA does not describe the physics of  Anderson
localization since, as discussed above, the arithmetically averaged
LDOS is non-critical at the Anderson transition.\cite{Wegner81}  To
overcome this deficiency Dobrosavljevi\'{c} and Kotliar
\cite{Dobrosavljevic97} formulated a variant of the DMFT where the
probability distributions (and not only the averages) of the local
Green functions are determined self-consistently (``Statistical
DMFT''). Employing a slave-boson mean-field theory as impurity solver
they investigated the disorder-driven MIT for infinitely strong
repulsion off half-filling. This statistical approach was also
employed in other studies of the Hubbard model\cite{atkinson} as well
as in the case of electrons coupled to phonons\cite{Fehske04} and the
Falicov-Kimball model.\cite{tran} Subsequently, Dobrosavljevi\'{c},
Pastor, and Nikoli\'c \cite{Dobrosavljevic03} incorporated the
geometrically averaged LDOS into the self--consistency cycle and
thereby derived  a mean--field theory of Anderson localization which
reproduces many of the expected features of the disorder--driven MIT
for non--interacting fermions. This scheme employs only one--particle
quantities and is therefore easily incorporated into the DMFT for
disordered electrons in the presence of phonons, \cite{Fehske04} or
Coulomb correlations.\cite{Byczuk05,Byczuk05a,Byczuk09,Aguiar09}

\subsection{Cold atoms in optical lattices: a new realization of disordered, correlated lattice quantum gases}

During the last few years cold atoms in optical lattices have emerged
as a unique tool--box for highly controlled investigations of quantum
many--body systems. In recent years the level of control in applying
disordered potentials to ultracold quantum gases has greatly improved
\cite{Aspect09,Lewenstein10}. Anderson localization in its pure form has been
demonstrated by the expansion of weakly interacting Bose-Einstein
condensates in a disordered speckle light field, giving rise to
characteristic localized condensate wave functions with exponentially
decaying tails \cite{Billy08,Roati08}. The additional influence of
strong repulsive interactions has been investigated recently in the
first full experimental realization of the 3d disordered Bose-Hubbard
model, by using a fine-grained optical speckle field superimposed by
an optical lattice \cite{White09}. In this experiment a strong
reversible suppression of the condensate fraction due to disorder was
observed, indicating the formation of a disorder-induced insulating
state. Independent experimental evidence was obtained from interacting
$^{87}$Rb bosons in a quasi-random (bichromatic) optical lattice,
where a strong reduction of the Mott gap was found and interpreted as
possible evidence for a compressible Bose glass phase
\cite{Fallani07}. On the theoretical side, low-dimensional
quasi-disordered Bose systems have been successfully described by DMRG
simulations \cite{Roux08}, which extended previous weak-coupling
calculations and found a direct transition from superfluid to Mott
insulator. Regarding disordered bosons in higher dimensions, the
status of theory is still more controversial, although significant
insight was gained by a new stochastic mean-field theory,
\cite{Bissbort09a} which allows for an efficient description of the
Bose glass phase and has already provided phase diagrams for realistic
speckle-type disorder \cite{Bissbort09b} such as used
experimentally \cite{White09}. Under debate remains the issue of a
direct transition between Mott insulator and superfluid, which was
claimed to be ruled out in recent QMC simulations in three spatial
dimensions, supported by general heuristic arguments
\cite{Pollet09}. Regarding disordered fermions, while no experiments
in cold gases have been performed yet, theory has significantly
advanced in recent years, mostly due to progress in the application of
DMFT to disordered and inhomogeneous systems
\cite{Dobrosavljevic97,Dobrosavljevic03,Byczuk05,Byczuk09,Semmler09}.
The phase diagram of spin-1/2 lattice fermions in a random potential
has now been determined theoretically, both in the paramagnetic
phase where Mott- and Anderson-insulator compete \cite{Byczuk05}, and
in the low-temperature regime where antiferromagnetic ordering sets in
and a new disorder-induced antiferromagnetic \emph{metallic} phase was
found \cite{Byczuk09}. In this way, also predictions for
single-particle spectral properties were obtained, which are now
becoming accessible experimentally via radio frequency
 spectroscopy measurements
of strongly interacting fermionic quantum gases \cite{Stewart08},
in analogy to photoemission  spectroscopy of electronic
solids. An alternative route towards single-particle spectroscopy
based on stimulated Raman transitions has been discussed theoretically
\cite{Dao07}.  Very recently, also the dynamical structure factor of
strongly interacting bosons in an optical lattice  has been measured
via two-photon Bragg scattering \cite{Ernst09,Fabbri09}.  These new
developments open the door towards controlled experimental realization
and spectroscopy of strongly interacting and disordered fermions in
optical lattices.

\subsection{Schematic phase diagram}

\begin{figure}[tpb]
\centerline{\includegraphics[scale = .4]{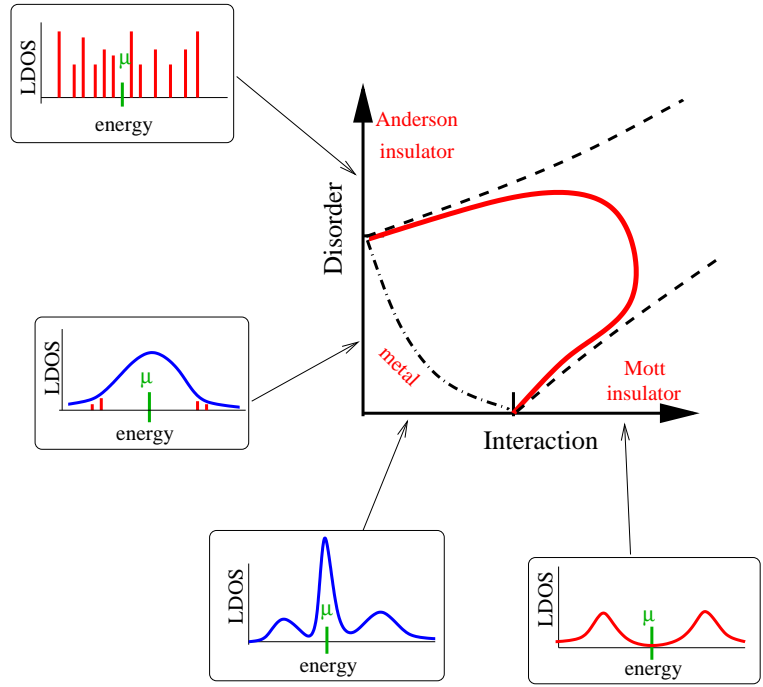}}
\caption{Schematic diagram of the possible phases and shapes of phase
  transition lines in disordered, interacting lattice fermion
  systems. In principle, the interplay between interactions and
  disorder could reduce the metallic regime (dash-dotted line), or
  enlarge it (full and dashed lines).  As will be discussed below,
  investigations within DMFT find that the metallic phase increases
  when interactions and disorder are simultaneously present (full
  line), and that the two insulating phases are   connected
  continuously, i.e., without critical behavior. Insets show the local
  density of states in the absence of disorder or interaction,
  respectively.}
\label{fig1.1}
\end{figure}

The Mott-Hubbard MIT is caused by short-range, repulsive interactions
in the pure system and is characterized by the opening of a gap in the
density of states  at the Fermi level. By contrast, the Anderson MIT
is due to the coherent backscattering of the quantum particles from
randomly distributed impurities in a system without interactions; at
the transition the character of the spectrum at the Fermi level
changes from a continuous to a dense point spectrum. Already these two
limits provide great challenges for theoretical investigations. It is
an even greater challenge to explore the \emph{simultaneous} presence
of interactions and disorder in lattice fermions systems. In view of
the construction of the dynamical mean-field approach employed here,
the results which will be presented in the following are expected to
provide a comprehensive description for systems in spatial dimensions
$d=3$ and larger, i.e., above the limiting dimension $d=2$. Two
particularly interesting questions are whether the metallic phase,
which exists at weak enough disorder and/or interaction strength, will
be reduced or enlarged, and whether the Mott and Anderson insulating
phases are separated by a metallic phase. Corresponding schematic
phase transition lines are shown in Fig.~\ref{fig1.1}. It is plausible
to assume that both MITs can be characterized by a single quantity,
namely, the local density of states. Although the LDOS is not an order
parameter associated with a symmetry breaking phase transition, it
discriminates between a metal and an insulator which is driven by
correlations and  disorder.

\section{Models of correlated fermions with disorder}

Here we study models of correlated fermions on ionic crystals or
optical lattices in the presence of diagonal (i.e., local) disorder
within a tight-binding description. In general, these models have the
form
\begin{equation}
H=\sum_{ij\sigma }t_{ij\sigma }c_{i\sigma }^{\dagger} c_{j\sigma}^{\phantom{+}}+ \sum_{i\sigma} \epsilon_i \; n_{i\sigma} + U\sum_{i}n_{i\uparrow }n_{i\downarrow },
\label{anderson_hubbard}
\end{equation}
where $c^{\dagger}_{i\sigma}$ and  $c_{i\sigma}$ are  the fermionic
creation and annihilation operators of the particle with spin
$\sigma=\pm 1/2$ at the lattice site $i$, $n_{i\sigma}=
c^{\dagger}_{i\sigma} c_{i\sigma}$ is the particle number operator
with eigenvalues $0$ or $1$, and $t_{ij\sigma }$ is the probability
amplitude for hopping between lattice sites $i$ and $j$. In the
Hubbard model $t_{ij\sigma}=t_{ij}$, i.e. the hopping amplitudes are
the same for both spin projections. In the Falicov-Kimball model
$t_{ij\sigma}=t_{ij}\delta_{\sigma \uparrow}$, i.e. only particles
with one spin projection are mobile and the others are localized. The
second term in (\ref{anderson_hubbard}) describes the additional
external potential $\epsilon_i$, which breaks the  ideal lattice
symmetry. For homogeneous systems we set $\epsilon_i=0$, which defines
the zero of the energy scale. The third term, a two-body term,
describes the increase of the energy by $U>0$ if two fermions with
opposite spins occupy the same site. In Eq.~(\ref{anderson_hubbard})
only  a local part of the Coulomb interaction is included and other
longer-range terms are neglected for simplicity. Note that this
approximation is excellent in the case of cold gases in optical
lattices, where the interaction between neutral atoms is essentially
local \cite{Jaksch98}. The disorder affects the system through a local
term $\sum_{i\sigma} \epsilon_i n_{i\sigma}$, where $\epsilon_i$ is a
random variable drawn from a probability distribution function (PDF)
${\cal P}(\epsilon_1,...,\epsilon_{N_L})$, where $N_L$ is a number of
lattice sites.  Typically we consider uncorrelated, quenched
  disorder, where
\begin{equation}
{\cal P}(\epsilon_1,...,\epsilon_{N_L})=\prod_{i=1}^{N_L}P(\epsilon_i)\,.
\end{equation}
Each of the $P(\epsilon_i)$ is the same, normalized PDF for the atomic
energies $\epsilon_i$. The quenched disorder means that
$P(\epsilon_i)$ is time independent. In other words, the atomic
energies $\epsilon_i$ are randomly distributed over the lattice and
cannot fluctuate in time. This type of disorder is different from
annealed  disorder where the random atomic energies have
thermal fluctuations.

In the following we use the continuous box-type PDF
\begin{equation}
P(\epsilon _{i})=\frac{1}{\Delta} \Theta (\frac{\Delta}{2}-|\epsilon
_{i}|)\,,
\label{four}
\end{equation}
with $\Theta (x) $  as the Heaviside step function. The parameter
$\Delta $ is therefore a measure of the disorder strength. The use of
a different continuous, normalized function for the PDF would bring
about only quantitative changes.

The Hubbard model and the Falicov-Kimball model defined by
(\ref{anderson_hubbard}) are not only of interest for solid-state
physics, but also in the case of  ultracold atoms, where specific
experimental realizations have been proposed. \cite{Lewenstein07} By
preparing a mixture of bosonic $^{87}$Rb and fermionic $^{40}$K in a
3d optical lattice, Ospelkaus {\it et al.} and G\"unter {\it et al.}
\cite{bose-fermi} were able to create -- to a first approximation -- a
version of the Falicov-Kimball model where the heavier bosonic species
could be slowed down even further by using a species-dependent optical
lattice and thus become ``immobile'' while the fermionic species
remains mobile. Alternatively, if the heavy bosonic species could be
frozen in a random configuration, this system would allow for a
realization of the Fermi-Hubbard model with quenched binary onsite
disorder. A different approach towards quenched randomness in optical
lattices was taken by White {\it et al.} \cite{White09} who
implemented a fine-grained optical speckle potential superimposed onto
a 3d optical lattice with interacting bosons and thus realized the
bosonic version of the Anderson-Hubbard model (\ref{anderson_hubbard})
with continuous disorder. A third alternative approach to disordered
cold gases is based on bichromatic optical lattices which are
quasiperiodic, as implemented for the 3d Bose-Hubbard model by Fallani
{\it et al.} \cite{Fallani07} who observed a disorder-induced
reduction of the Mott excitation gap, similar as discussed in the
following for the fermionic case.

The Hamiltonian~(\ref{anderson_hubbard}) is not solvable in
general. Without disorder, i.e., for $\Delta=0$, exact solutions on
an arbitrary lattice and in arbitrary dimension exist only for
$U=0$ (non-interacting fermions), or $t_{ij\sigma}=0$ (fermions in the
atomic limit). In the $U=0$ case the solution is obtained via discrete
Fourier transform, i.e.,
\begin{equation}
H=\sum_{{\bf k}\sigma} \epsilon_{{\bf k}\sigma} c_{{\bf k}\sigma}^{\dagger}c_{{\bf k}\sigma}^{\phantom{+}},
\end{equation}
where $\epsilon_{{\bf k}\sigma} = \sum_{j(i)} t_{ij\sigma} e^{-i{\bf
    k} ({\bf R}_j - {\bf R}_i)}$ are free fermion dispersion
relations. In the thermodynamic limit $N_L\rightarrow \infty$ the
spectrum is continuous and eigenstates are extended. In the
$t_{ij\sigma}=0$ limit the lattice sites are uncorrelated and the
  exact partition function has the form $ Z=\prod_i Z_i$, where
\begin{equation}
Z_i = 1+ 2 e^{\beta \mu }+  e^{-\beta U },
\end{equation}
where $\mu$ denotes the chemical potential within the grand canonical
ensemble, and $\beta =1/k_BT$ is the inverse temperature. In the
thermodynamic limit the spectrum is point-like and the eigenstates are
localized.

For finite disorder ($\Delta \neq 0$) an exact solution of the
Hamiltonian~(\ref{anderson_hubbard}) exists only for
$t_{ij\sigma}=0$. For a given realization of  disorder, i.e., when all
values of  $\left\{ \epsilon_1, \; \epsilon_2, \; ...., \;
  \epsilon_{N_L} \right\}$ are fixed, the partition function of the
model~(\ref{anderson_hubbard}) is given by
\begin{equation}
Z=\prod_iZ_i = \prod_i \left( 1+ 2 e^{-\beta(\epsilon_i- \mu) } + e^{-\beta U }\right).
\end{equation}
As in the atomic limit discussed above ($t_{ij\sigma}=0$) the spectrum is
point-like in the thermodynamic limit and the eigenstates are
localized.

The non-interacting limit ($U=0$) of (\ref{anderson_hubbard})  with
$t_{ij\sigma}\neq 0$ and disorder $\Delta \neq 0$ is not exactly
solvable. In a seminal paper by Abou-Chacra, Thouless,
Anderson\cite{ATA} the model (\ref{anderson_hubbard}) with $U=0$ and
$t_{ij\sigma}=t$ between nearest neighbor sites was solved on the
Bethe lattice, which is a tree-like graph without loops
\cite{Efetov97,Mirlin08}. The solution is expressed by the
one-particle Green function
\begin{equation}
G_{ii}(\omega) = \langle i | \frac{1}{\omega-H}| i\rangle =
\frac{1}{\omega-\epsilon_i - \eta_i(\omega)},
\end{equation}
where the hybridization function
\begin{equation}
\eta_i(\omega)=\sum_{j\neq i} \frac{t^2}{\omega-\epsilon_j - \eta_j(\omega)}
\end{equation}
describes a resonant coupling of site $i$ with its neighbors. If in
the thermodynamic limit the imaginary part of $\eta_i(z)$ is finite in
some band of energies $z$, then the states with energies $z$ are
extended. Otherwise, if the imaginary part of $\eta_i(z)$ is finite at
discrete energies $z$ such states are localized. For bound states
these energies $z$ form a point spectrum, and for Anderson localized
states  the energies $z$ form a dense point-like spectrum in the
thermodynamic limit. The analysis of the self-consistent equations
derived for $\eta_{i}(z)$ by Abou-Chacra, Thouless, Anderson\cite{ATA}
showed that, indeed, continuous and dense point spectra are separated
by a mobility edge which depends on the value of the disorder
$\Delta$.

In the following we solve the full
Hamiltonian~(\ref{anderson_hubbard}) by applying a dynamical
mean-field approximation to deal with the interaction and then discuss
how to cope with disorder.

\section{Dynamical mean-field theory}

The dynamical mean-field theory (DMFT) started from the following
observation\cite{Metzner89}: If the hopping amplitudes are scaled with
fractional powers of the space dimension $d$ (or the coordination
number $Z$), i.e., $t=t^*/\sqrt{2d}=t^*/\sqrt{Z}$ for nearest
neighbour hopping on a hypercubic lattice, then in the limit
$d\rightarrow \infty$ ($Z \rightarrow \infty$) the self-energy
$\Sigma_{ij}(\omega)$ in the Dyson equation
\begin{equation}
G_{ij\sigma}(i\omega_n)^{-1}= G_{ij\sigma}^0(i\omega_n)^{-1} - \Sigma_{ij\sigma}(i\omega_n),
\label{dyson}
\end{equation}
(here in a real-space representation) becomes diagonal \cite{MH89}
\begin{equation}
\Sigma_{ij\sigma}(i\omega_n)= \Sigma_{i\sigma}(i\omega_n)\;\delta_{ij},
\label{local}
\end{equation}
where $\omega_n=(2n+1)\pi/\beta$ are fermionic Matsubara
frequencies. In a homogeneous system the self-energy is site
independent, i.e.,
$\Sigma_{ij\sigma}(i\omega_n)=\Sigma_{\sigma}(i\omega_n)\;\delta_{ij}$,
and  is only a function of the energy. The DMFT approximation when
applied to finite dimensional systems neglects off-diagonal parts of
the self-energy. In other words, the DMFT takes into account all
temporal fluctuations but neglects spatial fluctuations between
different lattice sites.\cite{Georges96,PT}

Here we apply the DMFT to correlated fermion systems with
disorder. Within DMFT we map a lattice site onto a single impurity,
which is coupled to the dynamical mean-field bath. This coupling is
represented by the hybridization function $\eta_{i\sigma}(\omega)$,
which is determined self-consistently. The mapping is performed for
all $N_L$ lattice sites.

The partition function for a particular realization of disorder
$\left\{ \epsilon_1, \; \epsilon_2, \; ...., \; \epsilon_{N_L}
\right\}$ is now expressed  as a product of the partition functions
which are determined for each impurity (representing lattice sites),
i.e.,
\begin{equation}
Z=\prod_{i}Z_i = \prod_{i}\exp\left(  \sum_{\sigma\omega_n}  \ln[i\omega_n+\mu-\epsilon_i - \eta_{i \sigma}(\omega_n)-\Sigma_{i\sigma}(\omega_n) ] \right)\,.
\label{partition_bis_bis}
\end{equation}
The mean-field hybridization function $\eta_{i \sigma}(\omega_n)$ is
formally a site- and time- dependent one-particle potential. In the
interaction representation, the unitary time evolution due to this
potential is described by the local, time-dependent evolution operator\cite{Freericks-RMP,Freericks-book}
\begin{equation}
U[\eta_{i\sigma}]=T_{\tau} e^{-\int_0^{\beta}d\tau \int_0^{\beta}d\tau' c^{\dagger}_{i \sigma} (\tau) \eta_{i\sigma}(\tau-\tau') c_{i\sigma}(\tau') }\,,
\label{evolution}
\end{equation}
where $c_{i\sigma}(\tau)$ evolves according to the atomic part
$H_i^{\rm loc}$ of the Hamiltonian~(\ref{anderson_hubbard}) in
imaginary Matsubara time $\tau\in (0,\beta)$, and $T_{\tau}$ is the
time ordering operator. We write the  partition function
(\ref{partition_bis_bis}) as a trace over the operator
\begin{equation}
Z= Z [\eta_{i\sigma}]=\prod_{i=1}^{N_L}{\rm Tr}\left[ e^{-\beta
    (H_i^{\rm loc}-\mu N_i^{\rm loc} )}  U[\eta_{i\sigma}]\right]\,,
\label{dmft}
\end{equation}
where $N_i^{\rm loc}$ is the local particle number operator.

Eq.~(\ref{dmft}) allows us to determine the local one-particle Green
function $G_{ii\sigma}(\omega_n)$ for a given dynamical mean-field
$\eta_{i \sigma}(\omega_n)$. Indeed, the local Green function is
obtained by  taking a functional logarithmic derivative of the
partition function (\ref{dmft}) with respect to
$\eta_{i\sigma}(\omega_n)$, i.e.,
\begin{equation}
G_{ii\sigma}(\omega_n)=-\frac{\partial \ln Z[\eta_{i\sigma}]
}{\partial \eta_{i\sigma}(\omega_n)}\,.
    \label{local_green}
\end{equation}
Then we find the local Dyson equations
\begin{equation}
\Sigma_{i \sigma}(\omega_n)= i\omega_n+\mu-\epsilon_i-\eta_{i
  \sigma}(\omega_n) -\frac{1}{G_{i i \sigma}(\omega_n)}\,,
\label{local_dyson}
\end{equation}
for each $N_L$ lattice sites. For a single realization of disorder
$\left\{ \epsilon_1, \; \epsilon_2, \; ...., \; \epsilon_{N_L}
\right\}$, Eqs.~(\ref{dyson}, \ref{local}, \ref{dmft},
\ref{local_green}, and \ref{local_dyson}) constitute a closed set of
equations. A solution of this set represents an approximate solution
of the Hamiltonian~(\ref{anderson_hubbard}).

\section{Arithmetic \emph{vs.} geometric averaging}

A solution of Eqs.~(\ref{dyson}, \ref{local}, \ref{dmft},
\ref{local_green}, and \ref{local_dyson}) is very difficult to obtain
in practice. For each of the $N_L$ impurities we need to determine the
evolution operator (\ref{evolution})  exactly. Using rigorous methods
this can be done only for small $N_L$. However, Eqs.~(\ref{dyson},
\ref{local}, \ref{dmft}, \ref{local_green}, and \ref{local_dyson})
should be solved in the thermodynamic limit, $N_{L}\rightarrow
\infty$. This latter requirement might be overcome by performing finite
size scaling analysis. But such an analysis requires a large number of
lattice sites $N_L$ to reliably distinguish Anderson localized states
from those belonging to the continuum. Here one faces a typical
trade-off situation in computational physics. The computational
problem is greatly reduced when the local interaction in
(\ref{anderson_hubbard}) is factorized as in a Hartree-Fock
approximation, whereby genuine correlations are
eliminated.\cite{Logan93,Trivedi,Imada} Such approximate treatments
can nevertheless provide valuable hints about the existence of
particular phases. In our
investigation\cite{Byczuk05,Byczuk05a,Byczuk09} we employed the DMFT
to include all local correlations as will be discussed in the next
Section.

If one could solve the DMFT equations exactly, one would obtain a set
of local densities of states (LDOS)
\begin{equation}
A_{i\sigma}(\omega) = -\frac{1}{\pi} {\rm Im}
G_{ii\sigma}(\omega_n\rightarrow \omega + i0^+)\,,
\label{spectral}
\end{equation}
which are random quantities depending on the particular disorder
realization $\left\{ \epsilon_1, \; \epsilon_2, \; ...., \;
  \epsilon_{N_L} \right\}$. Usually one needs information about a
system that does not depend on a particular disorder
realization. Therefore one needs a statistical interpretation of the
solutions of Eqs.~(\ref{dyson}, \ref{local}, \ref{dmft},
\ref{local_green}, and \ref{local_dyson}).

When the system is large (cf., $N_L\rightarrow \infty$ in
thermodynamic limit) one usually takes the arithmetic average of the
LDOS   $A_{i\sigma}(\omega)$ over  many realizations of the disorder,
i.e.,
\begin{equation}
\langle A_{i\sigma}(\omega) \rangle= \int \prod_{j=1}^{N_L}d\epsilon_j\; P(\epsilon_i)  \; A_{i\sigma}(\omega; \{\epsilon_1,...,\epsilon_{N_L}\})\,,
\label{arith}
\end{equation}
where the dependence on $\left\{ \epsilon_1, \; \epsilon_2, \; ....,
  \; \epsilon_{N_L} \right\}$ is written explicitly. However, such a
method holds only if the system is self-averaging. This means that
sample-to-sample fluctuations
\begin{equation}
D_{N_L}(A_{i\sigma}(\omega))=\frac{\langle A_{i\sigma}(\omega)^2\rangle -\langle A_{i\sigma}(\omega)\rangle^2}{\langle A_{i\sigma}(\omega)\rangle^2}
\end{equation}
vanish for $N_L\rightarrow \infty$, which is equivalent to the Central
Limit Theorem for independent random variables
$A_{i\sigma}(\omega)$. By performing the arithmetic average one
restores the translational invariance in the description of the
disordered system, i.e., $A_{\sigma}(\omega)_{\rm arith} = \langle
A_{i\sigma}(\omega) \rangle$ is the same for all lattice sites.

An example of a \emph{non}-self-averaging system is a disordered
system at the Anderson localization transition, or a system whose
localization length is smaller than the diameter of the
sample.\cite{Anderson58} It implies that during the time evolution, a
particle  cannot explore the full phase space, i.e., cannot probe all
possible random distributions. In such a case the arithmetic average
(\ref{arith}) is inadequate. Here  one is faced with the question
concerning the proper statistical description of such a system.

The answer was given by Anderson:\cite{Anderson58} One should
investigate the full PDF for a given physical observable
$P[A_{i\sigma}(\omega)]$  and find  its most probable value, the
``typical'' value $A_{\sigma}(\omega)_{\rm typ}$, for which the PDF
$P[A_{i\sigma}(\omega)]$ has a global maximum. The typical value of
the LDOS, $A_{\sigma}(\omega)_{\rm typ}$, is the same for all lattice
sites. By employing $A_{\sigma}(\omega)_{\rm typ}$ one restores
translational invariance in the description of a disordered
system. This value will represent  typical properties of the
system. Using photoemission spectroscopy one could, in principle,
probe the LDOS at a particular lattice site and measure its most
probable value. We note that if sample-to-sample fluctuations are
small, the typical value $A_{\sigma}(\omega)_{\rm typ}$ would coincide
with the arithmetic average $A_{\sigma}(\omega)_{\rm arith}$. On the
other hand, in a non-self-averaging system the PDF can be strongly
asymmetric, with a long tail, in which case the typical value
$A_{\sigma}(\omega)_{\rm typ}$ would be very different from
$A_{\sigma}(\omega)_{\rm arith}$. The arithmetic mean is strongly
biased by rare fluctuations and hence  does not represent the typical
property of such a system.

Statistical approaches based on the computation of the probability
distribution functions would require the inclusion of very many
(perhaps infinitely many) impurity sites. This is very hard to achieve
in practice, in particular, in correlated electron systems discussed
here, although there have been recent successful attempts in this
direction\cite{Semmler09}. Therefore, one should look for a
generalized average which yields the best approximation to the typical
value. Among different means the \emph{geometric} mean turns out to be
very convenient to describe Anderson localization. The geometric mean
is defined by
\begin{equation}
A_{\sigma}(\omega) _{\mathrm{geom}}=\exp \left[ \langle \ln
  A_{i\sigma}(\omega)\rangle \right]\,,
\label{geom}
\end{equation}
where  $\langle F(\epsilon_i) \rangle =\int \prod_i d\epsilon
_{i}\mathcal{P}(\epsilon _{i})F(\epsilon _{i})$ is the arithmetic mean
of the function $F(\epsilon_i)$. The geometric mean is an
approximation to the most probable, typical value of the LDOS
\begin{equation}
A_{\sigma}(\omega)_{\rm typ}\approx A_{\sigma}(\omega) _{\mathrm{geom}}.
\end{equation}
It is easy to see that if $P[A_{i\sigma}(\omega)]$ is given by a
log-normal PDF then $A_{\sigma}(\omega)_{\rm typ} = A_{\sigma}(\omega)
_{\mathrm{geom}}$ holds exactly. It was shown that in the
non-interacting case  $A_{\sigma}(\omega) _{\mathrm{geom}}$ vanishes
at a critical strength of the disorder, hence providing  an explicit
criterion for Anderson localization
\cite{Anderson58,Dobrosavljevic97,Dobrosavljevic03,Schubert03}. We
also note that by using the geometrically averaged LDOS we restore the
translational invariance in our description of a disordered system. In
addition, as we shall see in the next Section, the restoration of
translational invariance by averaging allows us to solve the DMFT
equations in the thermodynamic limit. The problem of finite-size
effects is then automatically absent.

\section{DMFT self-consistency conditions for disordered systems}

According to the  spectral theorem the geometrically averaged local
Green function is given by
\begin{equation}
G_{\sigma} (\omega_n )_{\rm geom} = \int d\omega
\frac{ A_{\sigma} (\omega )_{\rm geom} }{i\omega_n-\omega }\,.
\label{local_green_average}
\end{equation}
The DMFT self-consistency condition (\ref{local_green}) is modified
now to a translationally invariant form
\begin{equation}
\Sigma_{\sigma}(\omega_n)= i\omega_n+\mu-\eta_{\sigma}(\omega_n)
-\frac{1}{  G_{\sigma}(\omega_n)_{\rm geom}}\,.
\label{self_invariant}
\end{equation}
Here we assumed that $\langle \epsilon_i\rangle = 0$, which holds in
particular for the box-shape PDF. We also used the translationally
invariant hybridization function $\eta_{\sigma}(\omega_n)$. We can now
perform a Fourier transform of the lattice Dyson equation
(\ref{dyson}) and obtain
\begin{equation}
G_{\sigma} (\omega_n )_{\rm geom} = \int d z \frac{N_0(z)}{i \omega_n -
  z +\mu  -  \Sigma_{\sigma}(\omega_n)},
\label{dyson_invariant}
\end{equation}
where $N_0(z)$ is the density of states for a non-interacting and
non-disordered lattice system.

Altogether the solution of the DMFT equations for interacting fermions
with disorder requires the following steps:
\begin{enumerate}
\item Select (i) $N_L$ values of $\epsilon_i$ from a given PDF
  $P(\epsilon_i)$, (ii) an initial hybridization function
  $\eta_{\sigma}(\omega_n)$, and (iii) an initial self-energy
  $\Sigma_{\sigma}(\omega_n)$;
\item for each $\epsilon_i$ solve the impurity problem defined by
  Eqs.~(\ref{evolution}, \ref{dmft}, \ref{local_green});
\item determine the LDOS $A_{i\sigma}(\omega)$ from the imaginary part
  of $G_{ii\sigma}(\omega)$, and  $A_{\sigma}(\omega)_{\mathrm{geom}}$
  from Eq.~(\ref{geom});
\item employ (\ref{local_green_average}) to find $G_{\sigma} (\omega_n
  )_{\rm geom}$;
\item from Eqs.~\ref{self_invariant} and \ref{dyson_invariant} find a
  new $\eta_{\sigma}(\omega_n)$ and $\Sigma_{\sigma}(\omega_n)$, then
  go to step (2) until convergence is reached.
\end{enumerate}
It is clear that  due to the averaging procedure we restore both
translational invariance and the thermodynamic limit although $N_L$ is
finite. Therefore the method is superior to other stochastic methods
which are affected by finite size effects.

In the presence of antiferromagnetic long-range order the
self-consistency conditions are modified. In this case we introduce
two sublattices $s=$A or B, and calculate two local Green functions
$G_{ii\sigma s}(\omega_n)$. From this quantity we obtain the
geometrically averaged LDOS  $A_{\sigma s}(\omega )_{\mathrm{geom}} =
\exp \left[ \langle \ln   A_{i\sigma s}(\omega)\rangle \right] $,
where $A_{i \sigma s}(\omega)$ is given as shown in
Eq.~(\ref{spectral}). The local Green function is then obtained from
the  Hilbert transform (\ref{local_green_average}). The local
self-energy $\Sigma _{\sigma s} (\omega )$ is determined from
Eq.~(\ref{self_invariant}). The self-consistent DMFT equations are
closed by the Hilbert transform of the Green function on a bipartite
lattice:
\begin{equation}
G_{\sigma s}(\omega_n )_{\mathrm{geom}} =\int dz \; \frac{N_{0}(z
  )}{\left[i \omega_n
-\Sigma_{\sigma s}(\omega_n )-\frac{z^2}{i \omega_n-\Sigma _{\sigma \bar{s}}(\omega_n)}\right]}.
\end{equation}
Here $\bar{s}$ denotes the sublattice opposite to $s$.\cite{Ulmke95,Georges96}

We note that if the geometric mean were replaced by the arithmetic
mean one would obtain a theory where disorder effects are described
only on the level of the CPA, which cannot detect Anderson
localization. It should also be pointed out that in the presence of
disorder the LDOS represented by $A_{\sigma}(\omega)_{\mathrm{geom}}$
is not normalized to unity. This means that
$A_{\sigma}(\omega)_{\mathrm{geom}}$ only describes the extended
states of the continuum part of the spectrum. Localized states, which
have a dense point spectrum, are not included in the DMFT with
geometric average. Therefore, this approach cannot describe the
properties of the Anderson-insulator phase.

The accuracy of the  DMFT approach with geometric average over
disorder was checked against numerically exact results obtained for
non-interacting fermions on a cubic
lattice.\cite{Dobrosavljevic03,Alvermann} The critical disorder
strengths at which Anderson localization occurs were found to agree
within a factor of two\cite{Alvermann} or
better\cite{Dobrosavljevic03}. However, there exists a discrepancy
regarding the shape of the mobility edge, which shows a pronounced
reentrant behavior for  non-interacting particles with box-type PDF of
the disorder. This feature  is not  reproduced by our
approach.\cite{Alvermann} On the other hand, the reentrant behavior
is a non-universal feature. Namely, it is much less pronounced in the
case of   a Gaussian PDF for disorder, and does not occur at all for a
Lorentzian PDF.\cite{Bulka}

It should be pointed out that the DMFT-based self-consistent approach
to interacting lattice fermions with disorder discussed here, is not
related to the self-consistent theory of  Anderson localization by
Vollhardt and W\"{o}lfle\cite{VW92} and its
generalizations.\cite{Nekrasov,Kroha} Namely, the   latter theory
determines  the frequency dependent diffusion coefficient $D(\omega)$
from arithmetically averaged two-particle correlation functions by
considering diffuson and cooperon diagrams. The approach reviewed here
does not make use of these coherent  back-scattering contributions,
but computes a one-particle correlation function, the LDOS, and
thereby extracts  information on Anderson localization. The fact that
the DMFT is based on a local approximation through the limit of large
spatial dimensions does not necessarily imply that back-scattering
contributions are entirely absent in this approach. Indeed,
contributions due to back-scattering are implicitly contained in the
hybridization function, which describes the diffusion of one-particle
excitations away from and back to a given lattice
site.\cite{Anderson58,ATA} Quite generally the relation between
theoretical approaches based on one-particle   and two-particle
correlation functions, respectively, and their results for the
critical disorder strength for Anderson localization, is still not
sufficiently understood and will continue to be an important topic for
future research. Perhaps the limit of high lattice dimensions will
serve as a useful starting point.\cite{Vlaming92,Janis01,Janis}

\section{Identification of different phases}

To characterize the ground state of the Hamiltonian
(\ref{anderson_hubbard}) the following quantities are computed:
\begin{enumerate}
\item the LDOS $A_{\sigma s}(\omega )_{\rm geom}$
for a given sublattice $s$ and spin direction $\sigma$;
\item the total DOS for a
given sublattice $s$ at the Fermi level ($\omega=0$) with  $N_s(0)_{\rm geom}\equiv
\sum_{\sigma}A_{\sigma s}(\omega=0)_{\rm geom}$;
\item the staggered
magnetization $m_{\mathrm{AF}}^{\rm geom}=|n_{\uparrow A}^{\rm geom}-n_{\uparrow
B}^{\rm geom}|$, where $n_{\sigma s}^{\rm geom}=\int_{-\infty}^0 d\omega A
_{\sigma s}(\omega )_{\rm geom}$ is the local particle density on
sublattice $s$.\cite{commenta}
\end{enumerate}
For comparison we determine these quantities also with the arithmetic
average.

The possible phases of the Anderson-Hubbard model can then be
classified as follows: The systems is a
\begin{itemize}
\item paramagnetic metal if $N_s^{\mathrm{geom}}(0)\neq 0$ and
  $m_{\mathrm{AF}}^{\mathrm{geom}}=0$;
\item  AF metal if $N_s^{\mathrm{geom}}(0)\neq 0$ and
  $m_{\mathrm{AF}}^{\mathrm{geom}}\neq 0$;
\item  AF insulator if $N_s^{\mathrm{geom}}(0)=0$ and
  $m_{\mathrm{AF}}^{\mathrm{geom}}\neq 0$ but $N_{
    s}^{\mathrm{geom}}(\omega )\neq 0$ for some $\omega \neq 0$
(in fact, the last condition is already implied by
$m_{\mathrm{AF}}^{\mathrm{geom}}\neq 0$);
\item paramagnetic Anderson-Mott insulator if
  $N_{s}^{\mathrm{geom}}(\omega )=0 $ for all $\omega$.
\end{itemize}
Note, that we use the term ``metal'' also for neutral fermionic atoms
if they fulfil the above conditions.

\section{Ground state phase diagram of interacting, disordered lattice fermion systems at half-filling}

We now apply the formalism discussed above to the Anderson-Hubbard
model at half-filling and compare the ground state properties in the
paramagnetic and magnetic cases.\cite{Byczuk05,Byczuk09}

 In the following we choose a model DOS, $N_{0}(\epsilon
 )=2\sqrt{D^{2}-\epsilon ^{2}}/\pi D^2$, with bandwidth $W=2D$, and
 set $W=1$. For this DOS and for a bipartite lattice the local Green
 function and the hybridization function are connected by the simple
 algebraic relation $\eta_{\sigma s}(\omega )_{\rm
   geom}=D^{2}G_{\sigma \bar{s}}(\omega )_{\rm
   geom}/4$. \cite{Georges96}

 The DMFT equations are solved at zero temperature by the numerical
 renormalization group technique \cite{NRG}, which allows us to
 calculate the geometric or arithmetic average of the local DOS in
 each iteration loop.

\subsection{Paramagnetic phase diagram}

\begin{figure}[tpb]
\centerline{\includegraphics[scale = .4]{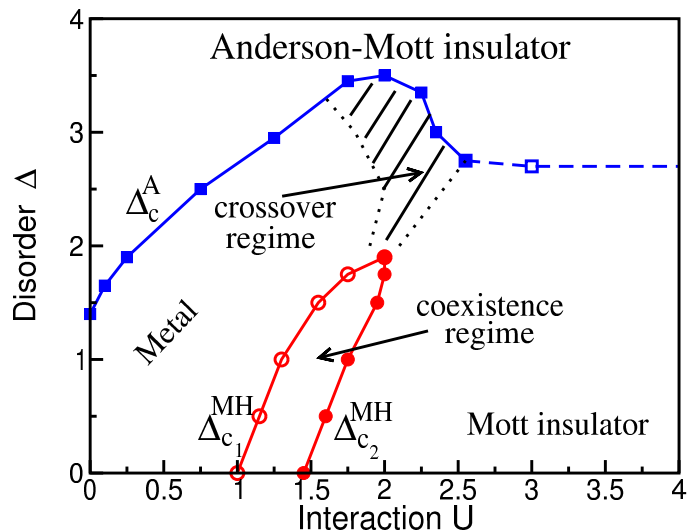}}
\caption{Non-magnetic ground state phase diagram of the
  Anderson--Hubbard model at half-filling as calculated by DMFT with
  the typical local density of states; after Ref.~58.}
\label{fig4.1}
\end{figure}

The ground state phase diagram of the Anderson-Hubbard model at
half-filling obtained within the DMFT approach discussed above is
shown in Fig. \ref{fig4.1}.\cite{Byczuk05} Two different phase
transitions are found to take place: a Mott-Hubbard MIT for weak
disorder $\Delta $, and an Anderson MIT for weak interaction $U$. The
correlated, disordered metal is surrounded by two different insulating
phases whose properties, as well as the transitions between them, will
now be discussed. In this section the spin index $\sigma$ is omitted
since all quantities are spin independent.

\emph{(i) Disordered, metallic phase}:The correlated, disordered metal
is characterized by a non-zero value of  the spectral density at the
Fermi level, $A(\omega=0)_{\rm geom}\neq 0$.  In the absence of
disorder DMFT predicts this quantity to be given by the bare DOS
$N_0(0)$, which is a consequence of the  Luttinger theorem.  This
means that Landau quasiparticles are well-defined at the  Fermi level.
The situation changes completely when disorder is introduced since a
subtle competition between disorder and  electron interaction arises.

\begin{figure}[tpb]
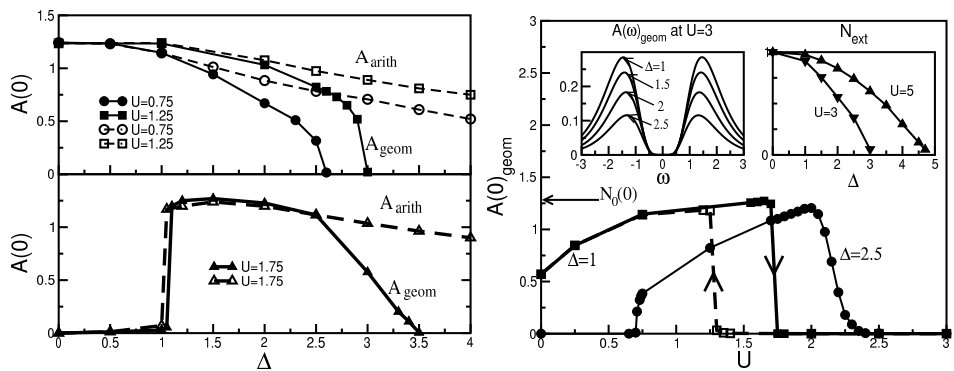

\centerline{\includegraphics[width=6.2cm]{Anderson_fig3a}
  \includegraphics[width=6.2cm]{Anderson_fig3b}}
\caption{Left panel: Local density of states (LDOS) as a function of disorder
  $\Delta$ for various values of the interaction $U$. Solid (dashed)
  curves correspond to the geometrically (arithmetically) averaged
  LDOS. Right panel: Geometrically averaged LDOS as a function of
  interaction $U$ for different disorder strengths $\Delta$. Solid
  (dashed) curves with closed (open) symbols are obtained with an
  initial metallic (insulating) hybridization function.  Triangles:
  $\Delta=1$; dots: $\Delta=2.5$. Left inset: LDOS with Mott gap at
  $U=3$ for different disorder strengths $\Delta$. Right inset:
  Integrated LDOS $N_{\rm{geom}}$ as a function of $\Delta$ at
  $U=3$; after Ref.~58.}
\label{fig4.2}
\end{figure}

Increasing the disorder strength at fixed $U$ reduces
$A(0)_{\rm{geom}}$ and thereby decreases the metallicity as shown in
the upper panel left of Fig. \ref{fig4.2}. The opposite behavior is
found  when  the interaction is  increased at  fixed $\Delta$ (see
right panel of Fig. \ref{fig4.2} for $\Delta=1$), i.e., in this case
the metallicity improves. In the strongly interacting metallic regime
the value of $A(0)_{\rm{geom}}$  is restored, reaching again its
maximal value $N_0(0)$.  This implies that in the metallic  phase
sufficiently  strong interactions protect the quasiparticles from
decaying by impurity scattering. For weak disorder this interaction
effect is almost independent of how the LDOS is averaged.

\emph{(ii)} \emph{Mott-Hubbard MIT}: For weak to intermediate disorder
strength there is a sharp transition at a critical value of $U$
between a correlated metal and a gapped Mott insulator. Two transition
lines are found depending on whether the MIT is  approached from the
metallic side [$\Delta^{MH} _{c2}(U)$, full dots in Fig. \ref{fig4.1}]
or from the insulating side [$\Delta^{MH} _{c1}(U)$, open dots in
Fig. \ref{fig4.1}]. The hysteresis is clearly seen  in right panel of
Fig. \ref{fig4.2} for $\Delta =1$.  The curves $\Delta^{MH}_{c1}(U)$
and $\Delta^{MH}_{c2}(U)$  in Fig. \ref{fig4.1} are seen to have
positive slope. This is a consequence of the disorder-induced increase
of spectral weight at the Fermi level which in turn requires  a
stronger interaction to open the correlation gap.  In the Mott
insulating phase close to the hysteretic region  an increase of
disorder will therefore drive the system \emph{back} into the metallic
phase. The corresponding abrupt rise of  $A(0)_{\rm{geom}}$ is clearly
seen in the left lower panel of Fig. \ref{fig4.2}. In this case the
disorder protects the metal from becoming  a Mott insulator.

Around $\Delta \approx 1.8$ the curves $\Delta^{MH}_{c1}(U)$ and
$\Delta^{MH}_{c2}(U)$ terminate at a single critical point,
cf. Fig. \ref{fig4.1}.  For stronger disorder ($\Delta \gtrsim 1.8$)
there appears to be a smooth crossover rather than a sharp transition
from the metal to the insulator. This is  illustrated by the $U$
dependence of $A(0)_{\rm{geom}}$ shown in right panel of
Fig. \ref{fig4.2} for $\Delta =2.5$. In this parameter regime the
Luttinger theorem is not obeyed for any $U$. In the crossover regime,
marked by the hatched area in Fig. \ref{fig4.1}, $A(0)_{\rm{geom}}$
vanishes gradually, so that the metallic and insulating phases can no
longer be distinguished rigorously \cite{bulla01}.

Qualitatively, we find that  the Mott-Hubbard MIT and the  crossover
region do not depend much on the choice of the average of the
LDOS.\cite{Byczuk05b} We also note the similarity between the
Mott-Hubbard MIT scenario discussed here for disordered systems and
that for a system without disorder at \emph{finite} temperatures
\cite{Georges96,bulla01},  especially the presence of a coexistence
region with hysteresis. However, while in the non-disordered case the
interaction needed to trigger the Mott-Hubbard MIT decreases with
increasing temperature, the opposite  holds in the disordered case.

\emph{(iii) Anderson MIT}: The metallic phase and the crossover regime
are found to lie next to an Anderson insulator phase where the LDOS
of the extended states vanishes completely (see Fig. \ref{fig4.1}).
The critical disorder strength $\Delta^A_{c}(U)$ corresponding to the
Anderson MIT  is a non-monotonous function of the interaction: it
increases in the metallic regime and  decreases  in the crossover
regime. Where $\Delta^A_c(U)$ has a positive slope an increase of the
interaction  turns the  Anderson insulator into a correlated metal.
This is illustrated in Fig. \ref{fig4.2} for $\Delta=2.5$: at
$U/W\approx 0.7$ a transition  from a localized to a metallic phase
occurs, i.e.,  the spectral weight at the Fermi level becomes finite.
In this case the electronic correlations inhibit the localization of
quasiparticles by scattering at the  impurities.

Fig. \ref{fig4.2} shows that the Anderson MIT is a continuous
transition. In the critical regime  $A(0)_{\rm{geom}}\sim \lbrack
\Delta^A_{c}(U)-\Delta]^{\beta }$ for $U=\rm{const}$.  In the
crossover regime a critical exponent $\beta =1$ is found (see the case
$U=1.75$ in lower panel of Fig. \ref{fig4.2}); elsewhere $\beta \neq
1$. However, we cannot rule out a very narrow critical regime with
$\beta =1$ since it is difficult to determine $\beta $ with high
accuracy. It should be stressed that an Anderson transition with
vanishing $A(0)_{\rm{geom}}$ at finite $\Delta =\Delta^A_{c}(U)$ can
only be detected within DMFT when the geometrically averaged LDOS is
used (solid lines in Fig. \ref{fig4.2}). Indeed, using arithmetic
averaging one finds a nonvanishing LDOS at any finite $\Delta $
(dashed lines in Fig. \ref{fig4.2}).

\emph{(iv) Mott and Anderson insulators}: The Mott insulator (with a
correlation gap) is rigorously defined only in the absence of disorder
($\Delta =0$), and the gapless Anderson insulator only for
non-interacting systems ($U=0$) and $\Delta >\Delta^A_{c}(0)$. For
finite interactions and disorder this distinction can no longer be
made. On the other hand, as long as the LDOS shows the characteristic
Hubbard subbands (left inset in Fig. \ref{fig4.2})  one may refer to a
\emph{disordered Mott insulator}. With increasing  disorder $\Delta $
the spectral weight of the Hubbard subbands vanishes (right inset in
Fig. \ref{fig4.2}) and the system becomes a \emph{correlated Anderson
  insulator}. The boundary between these two types of  insulators is
marked by a dashed line in Fig. \ref{fig4.1}. The results obtained
here within DMFT show that the paramagnetic Mott and Anderson
insulators are  continuously connected.  Hence, by changing $U$ and
$\Delta$ it is possible to move from one insulating state to  another
one without crossing a metallic phase.

\subsection{Magnetic phase diagram}

At half filling and in the absence of frustration effects interacting
fermions order antiferromagnetically. This raises several basic
questions: (i) How is a non-interacting, Anderson localized system at
half filling influenced by a local interaction between the particles?
(ii) How does an antiferromagnetic (AF) insulator at half filling
respond to disorder which in the absence of interactions would lead to
an Anderson localized state? (iii) Do Slater and Heisenberg
antiferromagnets behave differently in the presence of disorder? Here
we provide answers to these questions by calculating the zero
temperature, magnetic phase diagram of the disordered Hubbard model at
half filling using DMFT together with a geometric average over the
disorder and allowing for a spin-dependence of the DOS.\cite{Byczuk09}

\begin{figure}[tpb]
\centerline{\includegraphics[scale = .4]{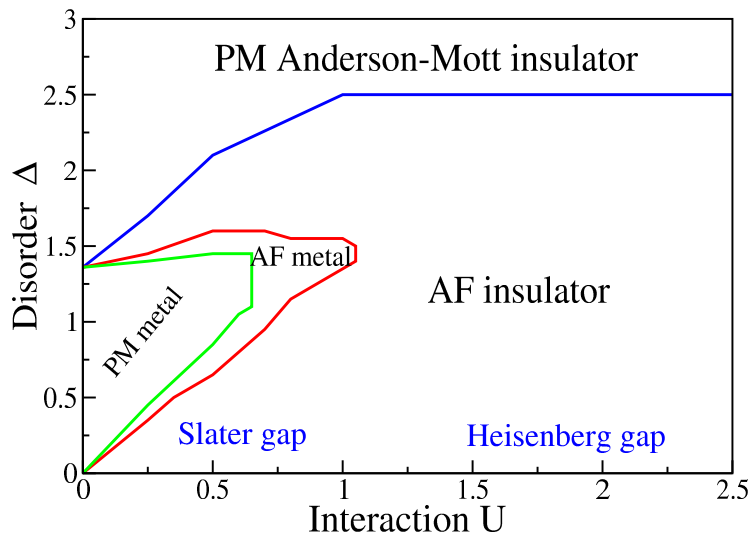}}
\caption{Magnetic ground state phase diagram of the Anderson-Hubbard
  model at half-filling as calculated by DMFT with a spin resolved
  local DOS (see text); PM: paramagnetic, AF: antiferromagnetic; after
  Ref. 60.}
\label{fig4.8}
\end{figure}

\begin{figure}[tpb]
\centerline{\includegraphics[scale = .4]{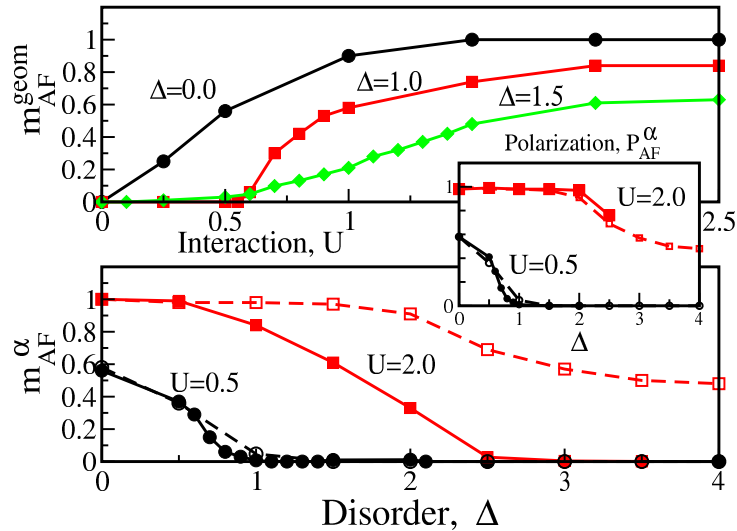}}
\caption{Upper panel: Staggered magnetization
  $m_{\mathrm{AF}}^{\mathrm{geom}}$ as a function of interaction
  $U$. Lower panel: $m_{\mathrm{AF}}^{\protect \alpha}$,
 $\protect\alpha =\mathrm{geom}/\mathrm{arith}$, as a function of
 disorder $\Delta $ (full lines: arithmetic average, dashed lines:
 geometric average). Inset: Polarization
 $P_{\mathrm{AF}}^{\protect\alpha }$ as a function of
 disorder.\cite{Byczuk09} Reprinted with permission from
 \it{Phys. Rev. Lett.} {\bf 102}, 146403 (2009). \copyright~ American Physical Society.}
\label{fig4.9}
\end{figure}

The ground state phase diagram of the Anderson-Hubbard model
(\ref{anderson_hubbard}) obtained by the above classification is shown
in Fig.~\ref{fig4.8}. Depending on whether the interaction $U$ is weak
or strong the response of the system to disorder is found to be very
different. In particular, at strong interactions, $U/W\gtrsim 1$,
there exist only two phases, an AF insulating phase at weak disorder,
$\Delta /W\lesssim 2.5$, and a paramagnetic Anderson-Mott insulator at
strong disorder, $\Delta /W\gtrsim 2.5$. The transition between these
two phases is continuous. Namely, the local DOS and the staggered
magnetization both decrease gradually as the disorder $\Delta $
increases and vanish at their mutual boundary (lower panel of
Fig.~\ref{fig4.9}). By contrast, the phase diagram for weak
interactions, $U/W\lesssim 1,$ has a much richer structure
(Fig.~\ref{fig4.8}). In particular, for weak disorder a
\textit{paramagnetic} metallic phase is stable. It is separated from
the AF insulating phase at large $U$ by a narrow region of \textit{AF}
\textit{metallic} phase. The AF metallic phase is long-range ordered,
but there is no gap since the disorder leads to a redistribution of
spectral weight.\cite{Byczuk09}

To better understand the nature of the AF phases in the phase diagram
we take a look at the staggered magnetization $m_{\mathrm{AF}}^{\alpha
}$. The dependence of $m_{\mathrm{AF}}^{\mathrm{geom}}$ on $U$ is
shown in the upper panel of Fig.~\ref{fig4.9} for several values of
the disorder $\Delta $. In contrast to the non-disordered case a
finite interaction strength $U>U_{c}(\Delta )$ is needed to stabilize
the AF long-range order when disorder is present. The staggered
magnetization saturates at large $U$ for both averages; the maximal
values depend on the disorder strength. In the lower panel of
Fig.~\ref{fig4.9} the dependence of $m_{\mathrm{AF}}^{\alpha }$ on the
disorder $\Delta$ is shown for different interactions $U$. Only for
small $U$ do the two averages yield approximately the same results.

Another useful quantity is the polarization  $P_{\mathrm{AF}}^{\alpha
}=m_{\mathrm{AF}}^{\alpha }/I^{\alpha }$, where $I^{\alpha
}=\int_{-\infty }^{+\infty }\sum_{\sigma s}\rho _{\sigma s}^{\alpha
}(\omega )d\omega /2$ is the total spectral weight of $\rho _{\sigma
  s}^{\alpha }(\omega )$. It allows one to investigate the
contribution of the point-like spectrum of the Anderson localized
states to the magnetization. This provides important information about
the spectrum since with increasing disorder more and more one-particle
states of the many-body system are transferred from the continuous to
the point-like spectrum. For weak interactions ($U=0.5$) the decrease
of the polarization with increasing disorder $\Delta $ obtained with
geometric or arithmetic averaging is the same (see inset in
Fig.~\ref{fig4.9}). Since within arithmetic averaging all states are
extended, the decrease of $m_{\mathrm{AF}}^{\alpha }$ (which is also
the same for the two averages in the limit of weak interactions, see
lower panel of Fig.~\ref{fig4.9}) must be attributed to disorder
effects involving only the continuous spectrum. At larger $U$ the
polarization is constant up to the transition from the AF insulator to
the paramagnetic Anderson-Mott insulator. In the latter phase the
polarization is undefined, because the continuous spectrum does not
contribute to $I_{\mathrm{AF}}^{\mathrm{geom}}$.

In the absence of disorder the AF insulating phase has a small
("Slater") gap at $U/W<1$ and a large ("Heisenberg") gap at
$U/W>1$. These limits can be described by perturbation expansions in
$U$ and $1/U$ around the symmetry broken state of the Hubbard and the
corresponding Heisenberg model, respectively. In agreement with
earlier studies \cite{Pruschke05} our results for $m_{\mathrm{AF}}$
(upper panel of Fig.~\ref{fig4.9}) show that there is no sharp
transition between these limits, even when disorder is present. This
may be attributed to the fact that both limits are described by the
same order parameter. However, the phase diagram (Fig.~\ref{fig4.8})
shows that the two limits \emph{can} be distinguished by their overall
response to disorder. Namely, the reentrance of the AF metallic phase
at $\Delta/W\gtrsim 1$ occurs only within the Slater AF insulating
phase.

The magnetic structure of the Anderson-Mott insulator cannot be
determined by the method used here since it describes only the
continuous part of the spectra and not the point spectrum. However,
only the paramagnetic solution should be expected to be stable because
the kinetic exchange interaction responsible for the formation of the
AF metal is suppressed by the disorder. This does not exclude the
possibility of Griffiths phase-like AF
domains.\cite{Griffiths,Miranda}

It is interesting to note that even the DMFT with an arithmetic average
finds a disordered AF metal \cite{Ulmke95,Singh98}. However, the
arithmetically averaged local DOS incorrectly predicts both the paramagnetic
metal and the AF metal to remain stable for arbitrarily strong disorder.
Only a computational method which is sensitive to Anderson localization,
such as the DMFT with geometrically averaged local DOS employed here, is
able to detect the suppression of the metallic phase for $\Delta /W\gtrsim
1.5$ and the appearance of the paramagnetic Anderson-Mott insulator at large
disorder $\Delta $ already on the one-particle level.

\section{Summary}

In this article we reviewed the properties of low-temperature quantum
phases of strongly correlated, disordered lattice fermion systems with
application to correlated electronic systems and ultracold fermions in
optical lattices. We discussed the Anderson-Hubbard model and a
comprehensive nonperturbative theoretical method for its solution, the
dynamical mean-field theory combined with geometrical averaging over
disorder. This approach provides a unified description of Anderson-
and Mott-localization in terms of one-particle correlation functions.

We presented low-temperature quantum phase diagrams for the
Anderson-Hubbard model at half filling, both in the paramagnetic and
the antiferromagnetic phase. In the paramagnet we observed re-entrant
metal-insulator transitions induced by disorder and interaction, where
the corresponding Anderson- and Mott-insulating phases are
continuously connected. In the presence of antiferromagnetic order, a
new antiferromagnetic metallic phase was found, which is stabilized by
the interplay between interaction and disorder.

It is expected that these new quantum states will be observable by
using ultracold fermions in optical lattices where disorder and
interactions are easily tunable in a wide range. While current
experimental temperatures are still above those required for observing
quantum antiferromagnetism, the paramagnetic Mott-Anderson insulator
should be easily accessible within current setups.

Even after several decades of research into the complex properties of disordered, interacting quantum many-body systems  many fundamental problems are still unsolved. Future investigations of the existing open questions, and of the new questions which are bound to arise, are therefore  expected to provide fascinating new insights.

\section*{Acknowledgements}
We thank R. Bulla and S. Kehrein for useful discussions. Financial support by the SFB 484, TTR 80, and FOR 801 of the Deutsche Forschungsgemeinschaft is gratefully acknowledged.

\printindex                         

\begin{thebibliography}{300}
\bibitem{definition1} The most probable value of a random quantity is
  defined as that value for which its PDF becomes maximal.
\end{thebibliography}
\end{document}